# Computational Neurology: Computational Modeling Approaches in Dementia

KongFatt Wong-Lin[1], Jose M. Sanchez-Bornot[1], Niamh McCombe[1], Daman Kaur[2], Paula L. McClean[2], Xin Zou[3], Vahab Youssofzadeh[4], Xuemei Ding[1], Magda Bucholc[1], Su Yang[1], Girijesh Prasad[1], Damien Coyle[1], Liam P. Maguire[1], Haiying Wang[5], Hui Wang[5], Nadim A.A. Atiya[1], Alok Joshi[1]

[1] Intelligent Systems Research Centre, School of Computing, Engineering and Intelligent Systems, Ulster University, Magee Campus, Northland Road, Derry~Londonderry, Northern Ireland, UK
[2] Northern Ireland Centre for Stratified Medicine, Biomedical Sciences Research Institute, Ulster University, Derry~Londonderry, Northern Ireland, United Kingdom
[3] Key Laboratory of Systems Biomedicine (Ministry of Education), Shanghai Centre for Systems Biomedicine, Shanghai Jiao Tong University, Shanghai, China
[4] Department of Neurology, Medical College of Wisconsin, Milwaukee, USA
[5] School of Computing, Ulster University, Jordanstown campus, Jordanstown, Northern Ireland, United Kingdom

Corresponding Author:
k.wong-lin@ulster.ac.uk (KongFatt Wong-Lin)
Intelligent Systems Research Centre, School of Computing, Engineering and Intelligent Systems, Ulster University, Magee Campus, Northland Road, Derry~Londonderry, BT48 7JL, Northern Ireland, UK

# Abstract

Dementia is a collection of symptoms associated with impaired cognition and impedes everyday normal functioning. Dementia, with Alzheimer's disease constituting its most common type, is highly complex in terms of etiology and pathophysiology. A more quantitative or computational attitude towards dementia research, or more generally in neurology, is becoming necessary – Computational Neurology. We provide a focused review of some computational approaches that have been developed and applied to the study of dementia, particularly Alzheimer's disease. Both mechanistic modeling and data-driven, including AI or machine learning, approaches are discussed. Linkage to clinical decision support systems for dementia diagnosis will also be discussed.

# Introduction

Neurology is a subdiscipline of medicine which seeks to understand and treat disorders of the nervous system. Many neurological or neurodegenerative diseases are often age-dependent, incurable or debilitating, while placing increasingly enormous socioeconomic burdens [45,23]. In particular, dementia is a collection of symptoms that is associated with impairment in cognition, particularly memory functioning, and impedes everyday normal functioning [20].

Alzheimer's disease (AD) is the most common form of dementia, and symptoms may involve memory and language impairment [20]. AD can be categorized under as familial AD (family history of the disease) and sporadic AD, with the latter overwhelmingly the most common type [17]. Various genes are currently thought to be associated with these different AD types. Mutations in amyloid precursor protein (APP), presenilin-1 (PSEN1) and presenilin-2 (PSEN2) are associated with familial AD while Apolipoprotein E (ApoE) gene has been linked to the sporadic type [20]. Other types of dementia include vascular dementia, frontotemporal dementia, Lewy body dementia, Huntington's disease, and Creutzfeldt-Jakob disease [20]. The intermediate stage between healthy and AD is labelled mild cognitive impairment (MCI) [20]. However, MCI is a loosely defined and heterogenous group, consisting of non-neurodegenerative or non-AD converters and people with other illnesses e.g. psychiatric illness [20]. Co-morbidities with other neurological disorders or illnesses, e.g. with epilepsy [42], are also well documented. With age as the primary risk factor, diagnosing and treating dementia and AD are now reaching a level of increasing urgency, not only because of the high proportion of cases relative to other neurodegenerative diseases, but also due to the rising global population and average lifespan [45,20].

To identify mechanisms and factors associated with AD, and its treatments, basic science and preclinical studies are necessary. Current preclinical models of AD pathologically associate the disease with beta-amyloid plaques (protein fragment snipped from an amyloid precursor protein) or tau neurofibrillary tangles (aggregates of hyperphosphorylated tau protein) formed in the brain in specific locations [20]. As the disease progresses, increasing evidence has shown that tau can spread from subcortical brain regions, particularly the entorhinal cortex and hippocampal region, to the rest of the brain regions via the connections between neurons i.e. synapses [9]. As tau is known to be toxic to synapses, it can cause damage to the connections as it propagates. A more recent model of AD associates it with neuroinflammation via abnormal glial cell activation [8].

Although there had been many potential AD drug developments in the pipeline, especially targeting beta-amyloids, they have mainly led to underwhelming results [32,11]. Currently US Food and Drug Administration (FDA) approved drugs such as acetylcholinesterase inhibitors are targeted to slow certain symptoms of cognitive decline, but not halting or curing AD [11]. A key challenge is the lack of surrogate biomarkers of AD for clinical studies. An example of surrogate markers for cardiovascular disease would be elevated cholesterol levels. Hence, the etiology and pathophysiology of AD remain to be completely characterized. Another challenge is the type of model. In particular, rodent models of AD, used in preclinical studies lack sufficient comparative research work linking them to human AD [18]. As we shall later discuss, computational models in the field of Computational Neuroscience may not only shed light on the complexity of the disease, and its co-morbidities, but also offer a bridge across species.

From a practical perspective, regardless of the etiology and pathophysiology, there is an urgent need to make informed clinical decisions for diagnosis and prognosis to allow appropriate care and treatment. For instance, AD or dementia is currently underdiagnosed (see e.g. [31]) and the misdiagnosis rates are seemingly high [21], as medical doctors or physicians in primary care may not be sufficiently or appropriately well trained to detect the disease within their limited consultation time

[6]. As we shall discuss below, data-driven computational, statistical or artificial intelligence (AI) models may help in not only mapping the relationships across symptoms and markers, but also act as clinical decision support systems by providing more objective and standardized information regarding AD status or progression.

Training or developing such computational or statistical models would require the availability of high quality and relevant attributes in datasets. However, in many clinical studies, cohorts are often small and selective. Datasets are also usually not openly available. Nevertheless, recent efforts and initiatives have moved towards larger scale studies, and big data integration and curation. Notable initiatives include the Alzheimer's Disease Neuroimaging Initiative (ADNI), the National Alzheimer's Coordinating Center (NACC), and the Dementias Platform UK (DPUK). Importantly, these open datasets now enable researchers, particularly, those with a computational or theoretical incline, to perform large-scale quantitative analyses to enable bigger research impact.

In the rest of this review, we shall focus on discussing how quantitative and computational approaches can be applied to address some of the aforementioned issues. We termed such approaches under the umbrella of Computational Neurology. We shall divide the discussions into three sections, focusing on AD, and each with a case study based on our previous work. The first section will discuss mechanistic modeling through Computational Neuroscience, and the second section will be on data-driven and AI approaches. Fig. 1 summarizes these computational approaches. The final section will provide a brief summary and the opportunities for further work.

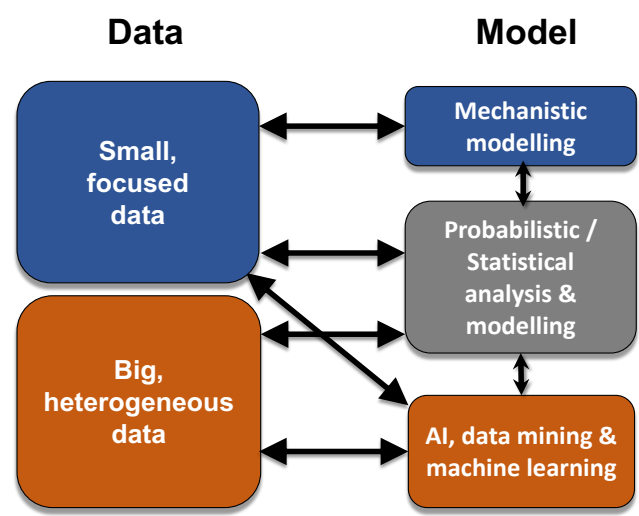

**Fig. 1. Schematic of computational and theoretical approaches in Computational Neurology.** Blue boxes: Small or focused data; brown: larger or more heterogeneous data. Arrows: Relationships. Sometimes AI, data mining and machine learning methods are also used in relatively smaller or less heterogeneous data (grey arrow) to guide mechanistic modeling (not shown).

## Computational Neuroscience: Mechanistic Modeling Approaches

If one is allowed to think of brain functions as some form of computational processing (though not necessarily processing like a computer, the understanding of brain functions can be divided into three levels of description (Marr, 1982) [33]. The first level is the computational theory, which deals with the goal of neural computation. This involves the appropriateness and logic of the strategy in the computation. The second level is the representations and algorithms of the processing. This entails how the computation is implemented or realized, what the inputs and outputs are, and the algorithms that process them. The third level is the "hardware" implementation, which describes how the representations and algorithms are represented in physical (e.g. biological) systems. The level to be employed depends on the research questions and also the availability of data types.

The field of Computational Neuroscience or Theoretical Neuroscience encompasses all these levels of description on brain functions [14]. In particular, the biophysical mechanistic understanding underlying neurodegeneration such as dementia can be enhanced with biophysically realistic or biologically motivated computational models [19,10]. Such models incorporate biological and physiological information of specific brain system(s) and apply mathematical modeling, especially differential equations to describe the systems' dynamics. Hence, the analyses of such models are often closely linked to dynamical systems theory [26,44]. Models often involve so-called conductance-based models, building on Hodgkin-Huxley type models [24], which make use of mathematical descriptions of ion-channel dynamics associated with neuronal excitability [28]. When the neurons in the models are "coupled" via mathematical description of their associated synaptic dynamics, neuronal circuit models can be simulated and studied. Hence, such models can, in principle, allow description across multiple levels, such as from ion channel dynamics through neuronal circuit dynamics to specific set of cognitive (e.g. memory) functions – multiscale modeling. It is the hope that cognition in normal and abnormal brains can be understood through modeling and simulation of neuronal circuit functions.

As an example, we have previously mentioned that the hippocampal brain region (Fig. 2A) is known to be one of the regions of deterioration in early-stage AD. The deterioration was associated with dysfunctions in memory, navigation and cognition, often observed in AD [20]. To investigate the possible mechanism of how beta-amyloid plaques can affect the activity and function of this region, a biophysical computational model of the hippocampal septal neuronal circuit model was modeled and simulated (Fig. 2B) [48,49].

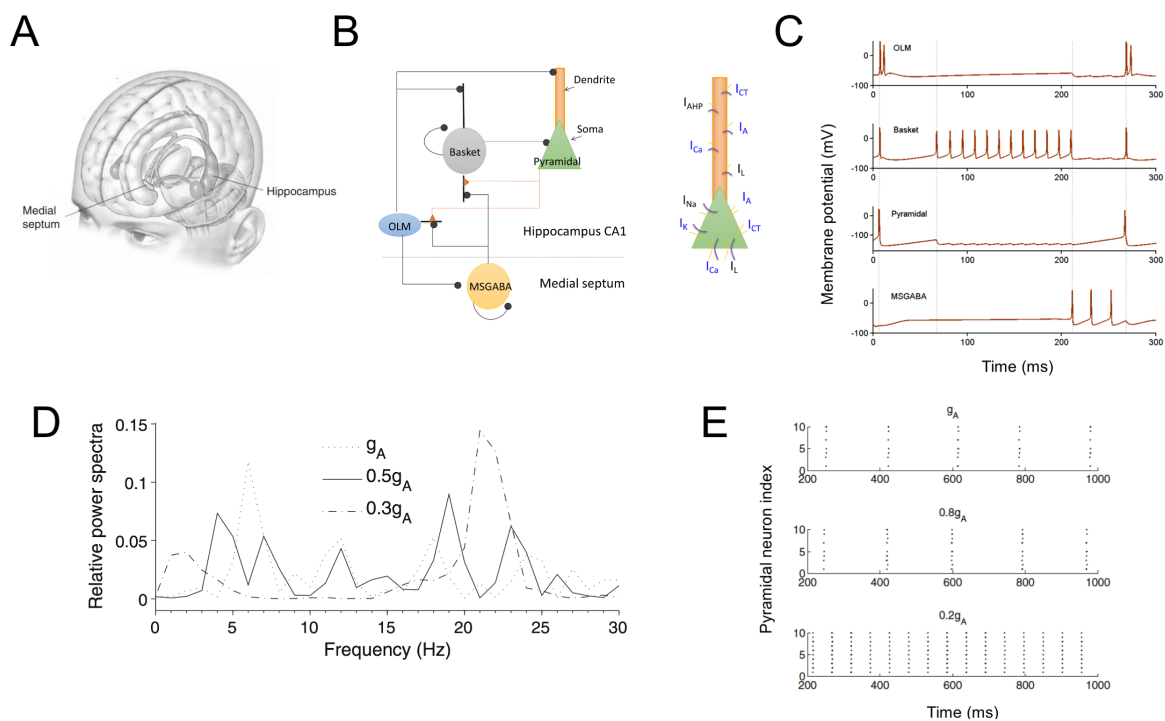

**Fig. 2. Neuronal circuit model and simulations. A.** Locations of the hippocampus and (medial) septum in the human brain. **B.** Neuronal circuit architecture of four populations of neurons in the hippocampal septal region. Excitatory (inhibitory) synapses: brown (black) filled circled-head arrows. Only the CA1 part of the hippocampus, and the medial part of the septum were modeled. **C.** Timecourse of neuronal membrane potential dynamics and spiking activities of a sample neuron from each neuronal population. Vertical dotted lines: Boundaries of rhythmic activity phases. **D.** Change in power spectra of neuronal population activity as conductance of A-type current, $g_A$, decreases. Higher frequency band peak increases while lower frequency band peak decreases. **E.** A hypothetical demonstration of neuronal hyperexcitability on memory deficits and epilepsy. Top: Spike raster gram of 10 modeled neurons. Correct memory pattern represented by pyramidal neurons 1, 3-5, 7, 9 and 10 using the original $g_A$ condition. Middle: Memory recall disrupted by low $g_A$ induced hyperexcitability due to additional neuronal spiking activity by the previously inactivated neurons (2 and 8). Bottom: Further hyperexcitation of pyramidal neurons lead to signs of epileptic seizure – comorbidity. A. Adopted with permission from [43]. B. Adapted with permission from [48]. D(right). C-E. Adapted with permission from [49].

In [48], a conductance-based Hodgkin-Huxley type neuronal model was developed to mimic the spiking activity of four types of hippocampal septal neurons, namely, the excitatory pyramidal neuronal types, and three inhibitory neuronal types (OLM, MSGABA, and Basket cells). The two-compartmental pyramidal neuronal membrane potential consists of a cell body called the soma, coupled to an extended part of the cell called dendrite in which other neurons can connect to excite/inhibit. The membrane potential dynamics were described by:

$$C_m \frac{dV_{dendrite}}{dt} = -I_L - I_{Ca} - I_{AHP} - I_A - I_{CT} - I_h - \frac{g_c}{1-p}(V_{dendrite} - V_{soma}) - I_{syn,dendrite}$$

$$C_m \frac{dV_{soma}}{dt} = -I_L - I_{Na} - I_K - I_{Ca} - I_A - I_{CT} - I_h - \frac{g_c}{p}(V_{soma} - V_{dendrite}) - I_{syn,soma} + I_e$$

where subscript *soma* and *dendrite* denoted soma and dendrite, respectively. $I_e$ was some injected DC current and $I_{syn}$'s were the summed synaptic currents. $g_c$ was the coupling strength between the soma and dendrite while $p$ was some fixed parameter. The rest of the $I$'s represent various ion-channel currents (with the types denoted as subscripts) based on the Hodgkin-Huxley type formalism with dynamics described by first-order kinetics (for details, see [48]). Many of the model parameters were constrained by neurophysiological data. For simplicity, the other three inhibitory neurons were modeled using one-compartment model (i.e. one differential equation describing each neuronal membrane dynamics), again consisting of various ion channel currents [48]. At baseline (control condition), the simulated neuronal spiking activities resembled those observed in experiments (Fig. 2c).

Based on several previous experimental studies, various ion-channel currents were altered in the model to indirectly simulate the effects of beta-amyloid plaques on the neuronal circuit. It was found that plaque-inhibited A-type fast-inactivating potassium ion channel current, $I_A$, can induce a significant effect on the neuronal population network dynamics [48]. Moreover, by analyzing the summed membrane potential dynamics, as $I_A$ decreases (via reduction in its conductance, $g_A$), lower frequency band power decreases while that of higher frequency band power increases (Fig. 2D) [48,49]. This was consistent with experimental observations (see e.g. [48,49] for some of the experimental findings described therein). The increase in oscillation band power was found to be due to the enhanced synchrony of the pyramidal neurons, while the decrease was due to the spiking phase relationship among the different neuronal types [49]. Interestingly, as the $I_A$ current was gradually inhibited (by beta-amyloid plaque), the hyperexcited pyramidal neurons first partially misremembered some earlier encoded information (cf. Figs. 2E (a) and (b)). Then with even higher hyperexcitation in the network, the latter exhibited sign of seizure (Fig. 2E (c)), as found in some AD patients – co-morbidity [42].

To summarize, mechanistic neuronal circuit modeling could help reveal specific downstream biological features (e.g. ion channels) affected by beta-amyloid plaques and provide a mechanistic explanation of the non-trivial changes in neuronal network dynamics. Such modeling could link mechanisms of co-morbidities (e.g. AD and epileptic seizures). Importantly, these models offer predictions that can be tested experimentally, e.g. between neuronal firing homeostasis and AD [1,41]. The models, usually described by differential/difference equations (or state maps), can be rigorously mathematically analyzed using applied dynamical systems theory (not shown, but see [49]). Moreover, models such as these can also (partially) address the issue of linking preclinical to clinical studies, with appropriate changes in the model parameters – the governing equations are largely similar.

Simpler (or more abstract) mechanistic models such as integrate-and-fire neuronal models [14,26,1] can also be employed if the neuronal excitability biophysical details (i.e. neuronal ion-channel dynamics) do not form part of the research questions. Similarly, population level (neural mass or mean-field) type models are also useful for AD investigation [4]. Other levels of modeling [33], especially at the computational cognitive level of description, include probabilistic, statistical, and information-theoretic types (see e.g. [14]) (Fig. 1). Again, these computational and theoretical tools are especially useful if biophysical details are not as important or readily available with respect to the research questions to be addressed.

## Data-Driven and AI Approaches

With the availability of neuroimaging technology and data, more direct analyses on the neurobiological progression of neurodegeneration in humans have become possible. For AD or dementia, (structural) magnetic resonance imaging (MRI), computed tomography (CT), and positron emission tomography (PET) scans are used as part of the clinical care pathway [34]. In academic communities, functional neuroimaging such as functional MRI (fMRI), electroencephalography (EEG), and more recently magnetoencephalography (MEG), are additionally used. The analyses of neuroimaging data are now providing a growing arena for the application of data-driven artificial intelligence (AI) techniques, and more recently, a subclass of AI called machine learning (ML) [46,2,13]. This is in parallel with (forward) model-driven analyses such as the neurobiologically based dynamic causal modeling (see e.g. [40,38]) or the more abstract oscillator-based modeling (e.g. [50]).

ML, applied to biomedical and healthcare data, mostly falls into the categories of supervised or unsupervised learning [36,37]. Supervised learning requires known labels or targets (e.g. disease classes), while unsupervised learning does not require such *a priori* knowledge. ML not only can provide more automation to analyses but may also offer higher classification accuracy of disease stage. For example, making use of their prowess in image processing, deep learning (deep neural network) models can remarkably now detect AD at around 96% accuracy and 84% for predicting conversion from MCI to AD [27].

ML can also make use of the integration of different neuroimaging modalities, taking advantage of their complementary strengths, to improve disease classification [39]. For example, using multi-kernel learning (MKL; for learning to identify an optimal kernel from a linear combination of kernels) with DARTEL (diffeomorphic anatomical registration through exponential Lie algebra) algorithm (for enhancing anatomical registration), [47] demonstrated that combined PET (with tracer PiB; Pittsburgh compound B) and MRI (grey matter, GM) data could enhance prediction of the three two-class classifications (using support vector machine): AD-vs-healthy (95.7%), MCI-vs-healthy (95.8%), and AD-vs-MCI (95.1%), as compared to using only PiB-PET (81.6%, 90.7%, 79.6%, respectively) or GM (91.3%, 92.5%, 89.1%, respectively). Similarly, using the leave-one-out cross-validation approach, a conservative and more clinically relevant approach, regression analyses on the multi-modal data for predicting individuals in AD, MCI or healthy classes provide a coefficient of determination of $r^2$=0.86, as compared to lower values using PiB-PET ($r^2$=0.61) or MRI ($r^2$=0.72) data alone (Fig. 3). The ML approach is sufficiently general to incorporate other non-neuroimaging data.

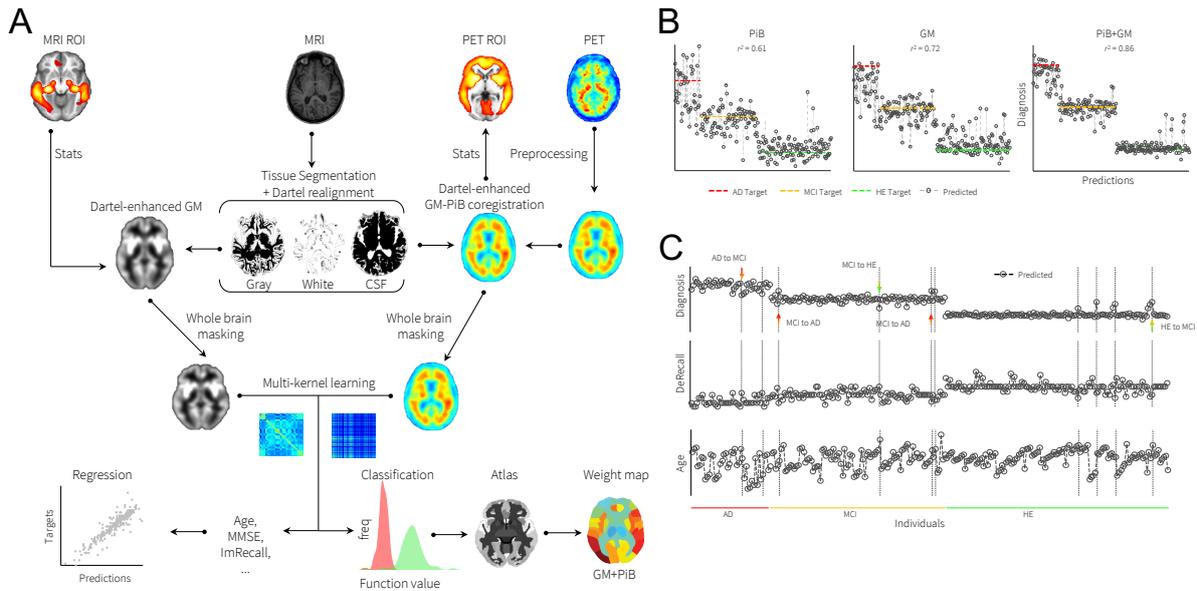

**Fig. 3. A machine learning framework based on multi-kernel learning (MKL) and DARTEL algorithms for predicting AD. A.** Framework that allows combination of multiple neuroimaging modalities (MRI, PET). **B,C.** A line prediction plot (predictions overlaid on targets) of diagnostic values of participants derived from: (GM-)MRI, (PiB-)PET scans, modeled by kernel ridge regression (KRR) method, and combined GM-MRI and PiB-PET data modeled by an optimized MKL technique, simpleMKL (or sMKL). Proximity of sample data to any horizontal line denotes the likelihood of classifying under that particular diagnostic category associated with that line. **B**. Group level; **C**. individual level with individual who transitioned across the classes (arrows). Adapted with permission from [47].

To date, several AI/ML approaches in AD research studies have focused on neuroimaging data. This is not too surprising, given the latter's complexity. However, other data types are gradually being investigated using ML. In clinical practice, most data consist of patients' electronic health/medical records such as medical history, family history, cognitive and functional assessments and sociodemographic, and typical blood-based tests. However, (typically structural) neuroimaging is conducted only if there is sufficient evidence of cognitive decline to warrant that, or to assist with distinguishing dementia subtypes (see e.g. [34,15]). Further, some of these non-imaging data can be readily collected longitudinally and in large quantities (big data). This would allow better understanding of the disease and its progression.

However, these data are usually highly heterogeneous (variety of data types) with multiple missing levels of description, e.g. data including specific genotype and memory recall performance. Hence, it is not practical or possible to develop mechanistic computational models based on such data. Hence, (big) data analytical approaches are more appropriate [25]. Often, these data have a large number of data features (i.e. variables, or sometimes called attributes or dimensions). These may require performing feature selection or dimensional reduction with e.g. supervised or unsupervised learning [22] to allow further in-depth analyses while reducing the risk of model overfitting. Such methods include information gain or recursive feature elimination types (e.g. [22,16,7]).

With the complexity of AD, data analytical approaches can holistically capture the relationship across multiple data features, for example, using probabilistic graphical modeling [30]. In particular, Bayesian network modeling, a specific type of directed acyclic graphical modeling of conditional dependences/probabilities (reflected as edges in a graph), is easy to use and interpret without too many assumptions [30,5]. For example, using an open dataset from the Australian Imaging Biomarkers and Lifestyle Study of Ageing (AIBL), [16] made use of Bayesian network modeling to capture the probabilistic causal relationship between very different data types as diverse as age, genotype, cognitive and functional assessments, and coarse-grained (total volume) neuroimaging data.

The Bayesian networks in [16] were generated using Hill Climbing score-based learning technique, in which network configurations with the highest score were selected [12]. Clinical dementia rating (CDR) score was used as the AD stage/class instead of using the more subjective clinical diagnosis. Interestingly, while following the same group of participants over two time points, it was found that the relationship can change over time, perhaps reflecting the dynamical changes due to ageing or the underlying disease (Figs. 4A-B). Notably, cerebrospinal fluid (CSF) total volume becomes more strongly influencing PET (with PiB tracer) and CDR, while grey matter (GM) total volume decreased in influence. This may suggest that these could be important biomarkers for older cohorts. The Bayesian network models can then be used to guide the combination of factors with predisposing factors and biomarkers GM, CSF, PiB-PET, ApoE, and age. Then a four-class (HC, very mild, mild, and moderate AD) classification was conducted. Various combinations of these data attributes were investigated, and some combinations obtained relatively high accuracy of identifying the AD class (Fig. 4C).

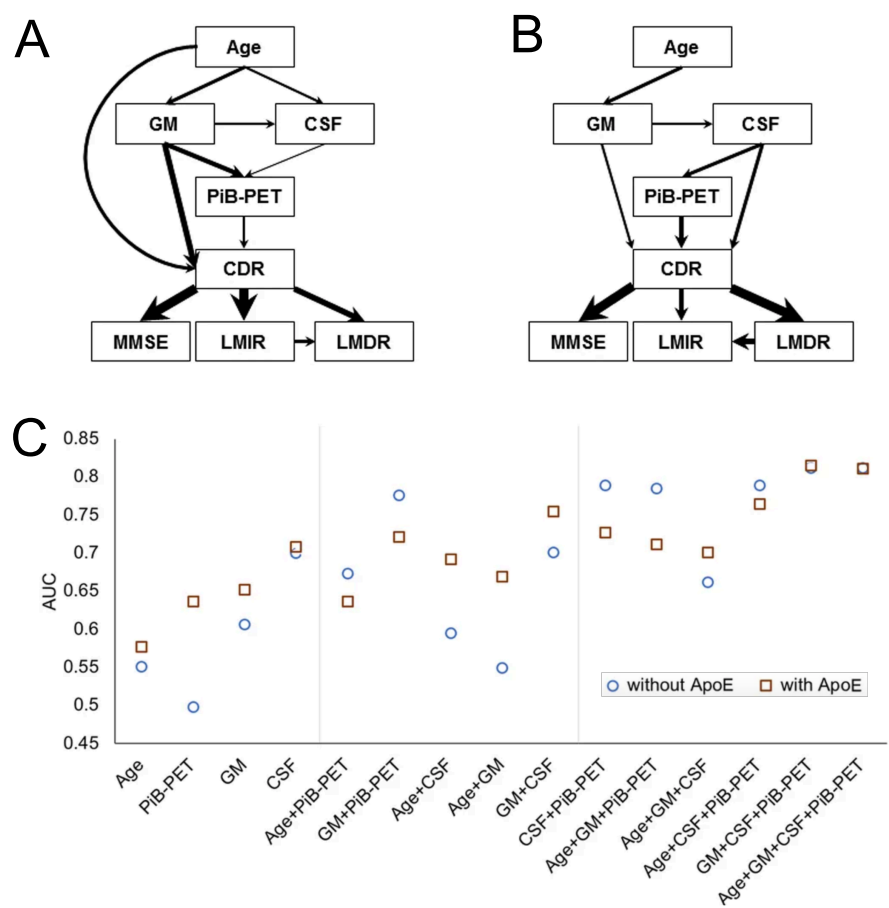

**Fig. 4. Bayesian network modeling identifies dynamical relationship and predisposing AD indicators and biomarkers. A,B.** Bayesian networks (BNs) generated based on 133 participants at baseline **(A)**, and at least one follow-up visit of the same participants during the M18-54 time interval **(B)**. Thickness of arrows: strength of probabilistic influences between variables. Only time-evolved features are included (e.g. (ApoE) genotype was excluded). Cognitive and functional assessments considered were: mini-mental state examination (MMSE), logical memory immediate/delayed recall assessments (LMIR/LMDR), and clinical dementia rating (CDR); neuroimaging features: structural MRI's total volume of grey matter (GM), cerebrospinal fluid (CSF), and positron-electron tomography (PET) with [$^{11}$C]-Pittsburgh compound B (PiB) tracer. **C.** Data consisted of all available complete samples regardless of time point. Blue circle (red square): without (with) ApoE genotype.

Classification accuracy (area under the receiver operating characteristic (ROC) curve, AUC) of individual predisposing indicators and biomarkers and their combinations with respect to CDR, ranked based on AUC. Circled markers: BN models constructed using individual as well as combinations of predisposing factors/biomarkers without ApoE. Squared markers: BN models constructed using individual as well as combinations of predisposing factors/biomarkers with ApoE. The incorporation of ApoE into the BN structure generally improved the model performance. Adapted with permission from [16].

With the availability of AI/ML-based, data-driven approaches, the route towards the development of automated clinical decision support systems (CDSS) [35] can be more readily laid out. For example, although most dementia diagnoses in clinical setting are based on categorical form, regression-based algorithms such as in [7] can provide CDSS with more graded information regarding disease stage (Fig. 5). This can not only provide more refined information regarding disease stages but is also more in line with dementia progression as a continuous dynamic process rather than discrete "jumps" across categories of severity.

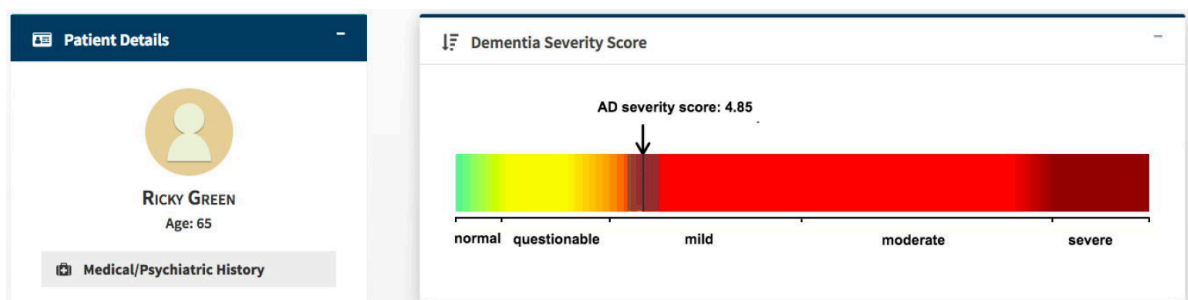

**Fig. 5.** Portion of the graphical user interface of a computer-based clinical decision support system for predicting dementia severity of an individual patient. Left: Patient information panel; right: AD severity measurement scale with AD severity score (black line) and its confidence interval (grey range). Adapted with permission from [7].

## Conclusion

We have discussed how neurological disorders such as neurodegeneration, particularly Alzheimer's disease, could be studied with guidance from several computational approaches, which we grouped under the umbrella term Computational Neurology. The computational approaches can generally be classified under model-based or data-driven/AI; the former is used if sufficiently detailed and focused datasets are available while the latter under the conditions in which the data are large and heterogeneous. Overall, there remain many opportunities for researchers to link these two computational approaches (e.g. see [3]).

The application of data-driven/AI in dementia research is still in its infancy, especially with regards to direct clinical applications. For the latter, sensitivity and specificity rates are especially important given the damaging consequences of misdiagnoses. Further, data-driven/AI approaches can readily lead to the development of automated clinical decision support systems. However, the latter are still not currently widely adopted in current dementia care pathways [29]. There are also challenges including their lack of pragmatism and realism for direct applications to frontline clinical settings. For example, some of the assessments in openly available datasets may not be used in clinical practice. Moreover, open datasets generally look "cleaner" than real-world data e.g. in electronic health/medical records. Perhaps deeper understanding of the needs and challenges of patients, clinicians and other stakeholders is needed to resolve these issues.

# Acknowledgments

The content was partially motivated by KWL's presentations at the Belfast Centre of Excellence meeting 2016 held at Queens University Belfast, the 7[th] Annual Translational Medicine (TMED) Conference, and the Biology of Brain Disorders (BBD) 2019: Frontiers in Innovation summer school. The authors, and some of the work discussed, were supported by: the CNRT award by the Northern Ireland Department for Employment and Learning through its "Strengthening the All-Island Research Base" initiative (XZ, GP, DC, LPM, KWL); Innovate UK (102161) (XD, MB, HYW, DHG, HW, GP, LPM, KWL); the Northern Ireland Functional Brain Mapping Facility (1303/101154803) funded by Invest NI and Ulster University (JMS-B, GP, LPM, DC, KWL); Northern Ireland International Health Analytics Centre (IHAC) collaborative network project funded by Invest NI through Northern Ireland Science Park (Catalyst Inc.) (VY, LPM, KWL); EU's INTERREG VA Programme, managed by the Special EU Programmes Body (SEUPB) (MB, LPM, DC, PLM, KWL); Ulster University Research Challenge Fund (MB, XD, PLM, KWL, JA); Global Challenges Research Fund (XD, MB, PLM, KWL, NAAA); ARUK NI Centre Pump Priming Awards (MB, JA, XD, PLM, KWL); COST Action Open Multiscale Systems Medicine (OpenMultiMed) supported by COST (European Cooperation in Science and Technology) (KWL); and the Dr. George Moore Endowment for Data Science at Ulster University (MB). The funders had no role in study design, data collection and analysis, decision to publish, or preparation of the manuscript, and the views and opinions expressed in this paper do not necessarily reflect those of the European Commission or the Special EU Programmes Body (SEUPB).

# References

1. K. Abuhassan, D. Coyle, A. Belatreche, L. Maguire (2014) Compensating for synaptic loss in Alzheimer's disease. J. Comput. Neurosci., 36(1), 19-37.
2. M.R. Ahmed, Y. Zhang, Z. Feng, B. Lo, O.T. Inan, H. Liao (2019) Neuroimaging and machine learning for dementia diagnosis: Recent advancements and future prospects. IEEE Rev. Biomed. Eng., 12, 19-33.
3. M. Alber, A. Buganza Tepole, W. R. Cannon, S. De, S. Dura-Bernal et al. (2019) Integrating machine learning and multiscale modeling—perspectives, challenges, and opportunities in the biological, biomedical, and behavioral sciences. npj Digit. Med., 2, 115. doi:10.1038/s41746-019-0193-y.
4. B.S. Bhattacharya, D. Coyle, L.P. Maguire (2011) A thalamo-cortico-thalamic neural mass model to study alpha rhythms in Alzheimer's disease. Neural Netw., 24(6), 631-645.
5. C.M. Bishop, *Pattern Recognition and Machine Learning*, Springer-Verlag, New York, 2007.
6. L. Boise, R. Camicioli, D.L. Morgan, J.H. Rose, L. Congleton (1999) Diagnosing dementia: Perspectives of primary care physicians. Gerontologist, 39(4), 457-464.
7. M. Bucholc, X. Ding, H. Wang, D.H. Glass, H. Wang et al. (2019) A practical computerized decision support system for predicting the severity of Alzheimer's disease of an individual. Expert. Syst. Appl., 130, 157-171.
8. V. Calsolaro, P.Edison (2016) Neuroinflammation in Alzheimer's disease: Current evidence and future directions. Alzheimer's Dement., 12(6), 719-732.
9. M. Costandi (2018) Ways to stop the spread of Alzheimer's disease. Nature, 559(7715), S16-S17.
10. D. Coyle, B.S. Bhattacharya, X. Zou, K. Wong-Lin, K. Abuhassan, L. Maguire, Neural circuit models and neuropathological oscillations, in: N. Kasabov (Ed.), *Springer Handbook of Bio-/Neuroinformatics*, Springer-Verlag, Berlin, Heidelberg, 2014.
11. J. Cummings, G. Lee, A. Ritter, K. Zhong (2018) Alzheimer's disease drug development pipeline: 2018. Alzheimer's Dement. (N.Y.), 4, 195-214.

12. R. Daly, Q. Shen, S. Aitken (2011) Learning Bayesian networks: approaches and issues. The Knowledge Engineering Review, 26(2), 99–157.
13. C. Davatzikos (2019) Machine learning in neuroimaging: Progress and challenges. Neuroimage, 197, 652-656.
14. P. Dayan, L.F. Abbott, *Theoretical Neuroscience*, The MIT Press, Cambridge, Massachusetts, London, England, 2001.
15. Dementia: Assessment, management and support for people living with dementia and their carers (2018) NICE guideline NG97.
16. X. Ding, M. Bucholc, H. Wang, D.H. Glass, H. Wang et al. (2018) A hybrid computational approach for efficient Alzheimer's disease classification based on heterogeneous data. Sci. Rep., 8(1), 9774.
17. J. Dorszewska, M. Prendecki, A. Oczkowska, M. Dezor, W. Kozubski (2016) Molecular basis of familial and sporadic Alzheimer's disease. Curr. Alzheimer's Res., 13(9), 952-963.
18. E. Drummond, T. Wisniewski (2017) Alzheimer's disease: Experimental models and reality. Acta Neuropathol., 133(2), 155-175.
19. L.H. Finkel (2000) Neuroengineering models of brain disease. Annu. Rev. Biomed. Eng., 2, 577-606.
20. S.A. Gale, D. Acar, K.R. Daffner (2018) Dementia. Am. J. Med. 131(10),1161-1169.
21. J.E. Gaugler, H. Ascher-Svanum, D.L. Roth, T. Fafowora, A. Siderowf, T.G. Beach (2013) Characteristics of patients misdiagnosed with Alzheimer's disease and their medication use: An analysis of the NACC-UDS database. BMC Geriatrics, 13, 137.
22. I. Guyon, A. Elisseeff (2003) An introduction to variable and feature selection. J. Mach. Learn. Res., 3, 1157-1182.
23. M. Heemels (2016) Neurodegenerative diseases. Nature 539, 179. doi:10.1038/539179a.
24. A.L. Hodgkin, A.F. Huxley (1952) A quantitative description of membrane current and its application to conduction and excitation in nerve. J. Physiol., 117(4), 500-544.
25. T. Hulsen, S.S. Jamuar, A.R. Moody, J.H. Karnes, O. Varga et al. (2019) From big data to precision medicine. Front. Med. (Lausanne). 2019; 6:34. doi:10.3389/fmed.2019.00034.
26. E.M. Izhikevich, *Dynamical Systems in Neuroscience: The Geometry of Excitability and Bursting*, The MIT Press, Cambridge, Massachusetts, London, England, 2007.
27. T. Jo, K. Nho, A.J. Saykin (2019) Deep learning in Alzheimer's disease: Diagnostic classification and prognostic prediction using neuroimaging data. Front. Aging Neurosci., 11, 20. doi:10.3389/fnagi.2019.00220.
28. D. Johnston, S.M.-S. Wu, *Foundations of Cellular Neurophysiology*, The MIT Press, Cambridge, Massachusetts, 1995.
29. S. Khairat, D. Marc, W. Crosby and A. A. S. Sanousi (2018) Reasons for physicians not adopting clinical decision support systems: Critical analysis. JMIR Med. Inform., vol. 6, no. 2, p. e24.
30. D. Koller, N. Friedman, *Probabilistic Graphical Models: Principles and Techniques*, The MIT Press, Cambridge, Massachusetts, London, England, 2009.
31. L. Lang, A. Clifford, L. Wei, D. Zhang, D. Leung et al. (2017) Prevalence and determinants of undetected dementia in the community: A systematic literature review and a meta-analysis. BMJ Open, 7(2), e011146. doi:10.1136/bmjopen-2016-011146.
32. S. Makin (2018) The amyloid hypothesis on trial. Nature, 559(7715), S4-S7.
33. D. Marr, *Vision: A Computational Approach*, San Francisco, Freeman & Co., 1982.
34. G.M. McKhann, D.S. Knopman, H. Chertkow, B.T. Hyman, C.R. Jack et al. (2011) The diagnosis of dementia due to Alzheimer's disease: Recommendations from the National Institute on Aging-Alzheimer's Association workgroups on diagnostic guidelines for Alzheimer's disease. Alzheimers Dement., 7(3), 263-269.
35. B. Middleton, D.F. Sittig, A. Wright (2016) Clinical decision support: A 25 year retrospective and a 25 year vision. Yearb. Med. Inform. (Suppl 1), S103–16.
36. T.M. Mitchell, *Machine Learning*, McGraw Hill Education, 2017.

37. P. Norvig, S. Russell, *Artificial Intelligence: A Modern Approach*, Pearson Education India, 3rd Ed., 2015.
38. W. Penny, J. Iglesias-Fuster, Y.T. Quiroz, F.J. Lopera, M.A. Bobes (2018) Dynamic causal modeling of preclinical autosomal-dominant Alzheimer's disease. J. Alzheimers Dis., 65(3), 697-711.
39. S. Rathmore, M. Habes, M.A. Iftikhar, A. Shacklett, C. Davatzikos (2017) A review on neuroimaging-based classification studies and associated feature extraxtion methods for Alzheimer's disease and its prodromal stages. Neuroimage, 155, 530-548.
40. R. Rystar, E. Fornari, R.S. Frackowiak, J.A. Ghika, M.G. Knyazeva (2011) Inhibition in early Alzheimer's disease: An fMRI-based study of effective connectivity. Neuroimage, 57(3), 1131-1139.
41. B. Styr, I. Slutsky (2018) Imbalance between firing homeostasis and synaptic plasticity drives early-phase Alzheimer's disease. Nat. Neurosci., 21(4), 463-473.
42. K.A. Vossel, M.C. Tartaglia, H.B. Nygaard, A.Z. Zeman, B.L. Miller (2017) Epileptic activity in Alzheimer's diseases: Causes and clinical relevance. Lancet Neurol., 16(4), 311-322.
43. A.M. White (2003) What Happened? Alcohol, memory blackouts, and the brain. Alcohol Res. Health, 27(2), 186-196.
44. H.R. Wilson, *Spikes, Decisions, and Actions: The Dynamical Foundations of Neuroscience*, Oxford University Press, Inc., New York, 1999.
45. World Health Organization, *Neurological disorders: public health challenges*. Geneva, World Health Organization, 2006.
46. S. Yang, J. Sanchez-Bornot, K. Wong-Lin, G. Prasad (2019) M/EEG-based bio-markers to predict the MCI and Alzheimer's disease: A review from the ML perspective. IEEE Trans. Biomed. Eng., 66(10), 2924-2935.
47. V. Youssofzadeh, B. McGuinness, L.P. Maguire, K. Wong-Lin (2017) Multi-kernel learning with dartel improves combined MRI-PET classification of Alzheimer's disease in AIBL data: Group and individual analyses. Front. Hum. Neurosci., 11, 380.
48. X. Zou, D. Coyle, K. Wong-Lin, L. Maguire (2011) Computational study of hippocampal-septal theta rhythm changes due to ß-amyloid-altered ionic channels. PLoS One, 6(6), e21579.
49. X. Zou, D. Coyle, K. Wong-Lin, L. Maguire (2012) Beta-amyloid induced changes in A-type K+ current can alter hippocampo-septal network dynamics. J. Comput. Neurosci., 32(3), 465-477.
50. T.H. Alderson, A.L.W. Bokde, J.A.S. Kelso, L. Maguire and D. Coyle (2018) Metasrable neural dynamics in Alzheimer's disease are disrupted by lesions to the structural connectome. Neuroimage, 183, 438-455.